# Lie detection algorithms attract few users but vastly increase accusation rates

[1,2]* Alicia von Schenk, [1,2]*Victor Klockmann, [3,4]Jean-François Bonnefon,

[2]Iyad Rahwan & [2]Nils Köbis

* shared first authorship

1 - Department of Economics, University of Würzburg

2 - Max Planck Institute for Human Development, Center for Humans and Machines

3 - Institute for Advanced Study Toulouse

4 - Toulouse School of Economics

**Correspondence to:** Nils Köbis, koebis@mpib-berlin.mpg.de

**Acknowledgements:** JFB acknowledges support from grant ANR-19-PI3A-0004, grant ANR-17-EURE-0010, and the research foundation TSE-Partnership.

## Abstract

People are not very good at detecting lies, which may explain why they refrain from accusing others of lying, given the social costs attached to false accusations — both for the accuser and the accused. Here we consider how this social balance might be disrupted by the availability of lie-detection algorithms powered by Artificial Intelligence. Will people elect to use lie detection algorithms that perform better than humans, and if so, will they show less restraint in their accusations? We built a machine learning classifier whose accuracy (67%) was significantly better than human accuracy (50%) in a lie-detection task and conducted an incentivized lie-detection experiment in which we measured participants' propensity to use the algorithm, as well as the impact of that use on accusation rates. We find that the few people (33%) who elect to use the algorithm drastically increase their accusation rates (from 25% in the baseline condition up to 86% when the algorithm flags a statement as a lie). They make more false accusations (18*pp* increase), but at the same time, the probability of a lie remaining undetected is much lower in this group (36*pp* decrease). We consider individual motivations for using lie detection algorithms and the social implications of these algorithms.

## Introduction

People lie a lot, in many contexts (DePaulo et al. 1996, Pascual-Ezama et al. 2020, Serota et al. 2010, Tergiman and Villeval 2022). In many contexts, it would be advantageous to detect lies and call them out (van den Assem et al. 2012, Turmunkh et al. 2019, Warren and Schweitzer 2018). While some methods help with lie detection (Nahari et al. 2014, Verschuere et al. 2022), the time, effort, and skill they require place them beyond the reach of ordinary people. Accordingly, recent studies (Pascual-Ezama et al. 2021) and large-scale meta-analyses indicate that people do not perform much better than chance when trying to detect lies (Hartwig and Bond 2011, Hauch et al. 2016). This general poor performance in lie detection may explain why people typically refrain from accusing others of lying (Gilbert 1991, Levine et al. 1999). Indeed, poor performance at lie detection increases the risk of making false accusations, which are harmful both to the accused and to the accuser. False accusations can be harmful to the accused because of the social stigma of being called a liar; and they can, in turn, be harmful to the accuser, who is held accountable for unjustly tarnishing the reputation of the accused. Since people are bad at detecting lies, it may be a safer strategy to refrain from accusations that can hurt both the accuser and the accused if they are unfounded.

As a corollary, anything that would reduce either the harm to the accused or the accountability of the accuser may upend our current social balance and increase the rate at which people accuse each other of lying. For example, the harm of false accusations to the accused can be reduced by systematic fact-checking. Currently, this time-consuming process is mostly reserved for high-stakes accusations (e.g., in judicial or political contexts) and is unlikely to be available in all accusation contexts. Technology may change that if fact-checking can be automated and scaled up; but the real technological game-changer may consist of automatic lie detection that decreases the accountability of the accuser rather than automated fact-checking that reduces harm to the accused.

Indeed, progress in Artificial Intelligence is opening a new chapter in the long history of lie-detecting machines. While older machines such as the polygraph have questionable accuracy (Saxe et al. 1985), current Natural Language Processing algorithms can detect fake reviews (Pérez-Rosas et al. 2017) and achieve higher-than-chance accuracy for lie detection (Kleinberg and Verschuere 2021). If this technology continues to improve and becomes massively available, it may disrupt the current social balance in which people largely refrain from accusing each other of lying.

Imagine a world in which everyone has access to a superhuman lie-detection technology, such as Internet browsers that screen social media posts for lies; algorithms that check CVs for deception; or video conferencing platforms that give real-time warnings when one's interlocutor or negotiation partner seems to be insincere (Gaspar and Schweitzer 2013). Consulting a lie detection algorithm, or delegating accusations to the algorithm, could reduce accusers' sense of accountability, increase the psychological distance from the accused, and blur questions of liability (Hohenstein and Jung 2020, Köbis, Bonnefon, et al. 2021), resulting in higher accusation rates.

This assumes, however, that people do elect to use such AI tools for lie detection. We know that people are often reluctant to use algorithms, especially when the algorithms are not error-proof (Dietvorst et al. 2018), and that this reluctance is especially high in emotion-laden domains (Castelo et al. 2019). As a result, the disruptive potential of lie detection algorithms may be neutralized or delayed by low adoption. In this work, we develop a lie-detection algorithm whose accuracy is better than that of humans, and we conduct an incentivized lie-detection experiment in which we measure participants' propensity to use the algorithm, as well as the impact of that use on accusation rates. We further measure the individual predictors of algorithmic uptake and show that they depend on the concurrent availability of fact-checking mechanisms.

## Methods

**Statement collection.** As preparation for our main study, we collected a dataset of true and false statements to be used for algorithm training and our lie-detection task. This dataset was collected in January 2022, and the main study took place in April 2022. All data collections were approved by the Ethics Committee of the *blinded for review.* Participants provided informed consent at the start of the study. Data sets and STATA analysis scripts are openly available via the Open Science Framework:

(https://osf.io/eb59s/?view_only=bf8c8f966c084941a59b117126e4aea8)

We recruited 986 participants and asked them to describe something they intended to do during the next weekend. While some studies let people decide whether to lie or not (Erat and Gneezy 2012, Gneezy 2005, Leib et al. 2021), we adopted standard procedures in research on lie detection and eliciting true and false statements from each participant (Kleinberg and Verschuere 2021, Verschuere et al. 2018). Participants were first required to write a true statement and then to write a false statement and incentivized to write convincingly (they earned a bonus of £2 if a future participant judged their statement to be true, see details below). They were not informed beforehand that they would have to write a false statement after the truthful one.

This approach has the advantage of obtaining better training data for the lie detection algorithm because (a) we avoid selection bias of lies stemming from the endogenous choice by participants, and (b) true and false statements are perfectly balanced in the training dataset. After writing the two statements, participants could pay a fee of £0.30 to prevent future participants from obtaining aid from lie-detection algorithms to judge their statements.

Two research assistants coded the quality of these statements. First, they checked whether the participant followed the instructions and wrote meaningful sentences with at least 150 characters. Second, for truthful statements, they checked whether the author's supporting text fitted the statement (see SI). Participants were excluded from the dataset if either of their statements did not meet either of these criteria. After this quality check, our dataset contained 1536 statements from 768 authors.

**Lie-detection algorithm.** To generate a lie-detection algorithm that performed better than humans, we relied on the open-source BERT language model by Google (Devlin et al. 2018). To obtain an out-of-sample prediction of truthfulness for each statement, we repeatedly performed an 80:20 split of the data. Specifically, we first split the 1536 statements into five equally sized subsets. We then trained the model with 80% of the data and tested it with the remaining 20%, and repeated this step five times. Accuracy was then measured by the total performance of the algorithm across all five folds. The algorithm reached 67% accuracy (=correctly identifying lies as lies and truthful statements as truthful), which was deemed sufficient to proceed to the next stage, given the expectation that humans would not perform much better than chance level at this task.

**Main study.** In all experimental treatments, participants ($N = 384$) read a total of four statements and decided each time whether to judge the statement as truthful or whether to accuse the author of lying. They were incentivized for accuracy (bonus of £0.50 for each correct guess) and were informed beforehand that half of the statements were truthful and the other half were lies. They also learned that the four statements they would see were picked randomly and that hence the actual number of truthful and lied statements we showed them was determined at random.

Participants were randomly assigned to four treatments that orthogonally manipulated the availability of a lie-detection algorithm and the presence of fact-checking. When the lie-detection algorithm was available, participants could purchase its prediction ("true statement" or "false statement") for a small cost of £0.05 (this cost had to be paid separately for each statement). If a participant purchased an algorithmic prediction about a statement whose author opted to pay for privacy, they were simply informed that the prediction was unavailable (without further elaboration), and their purchase was refunded.

Orthogonally to the availability of the lie-detection algorithm, we manipulated the presence of a fact-checking mechanism. When the fact-checking mechanism was absent, participants were informed that whenever they accused an author of lying, the payoff of this author would decrease by £2.00 (or 40% of the author's total achievable payoff). When the fact-checking mechanism was present, participants were informed that when they accused an author of lying, the payoff of this author would decrease by £2.00 only if the accusation was correct. This allowed us to compare the effects of fact-checking and lie-detection algorithms on accusation rates, as well as to explore their combined effect. Hence, we had a total of four treatments (see Figure 1).

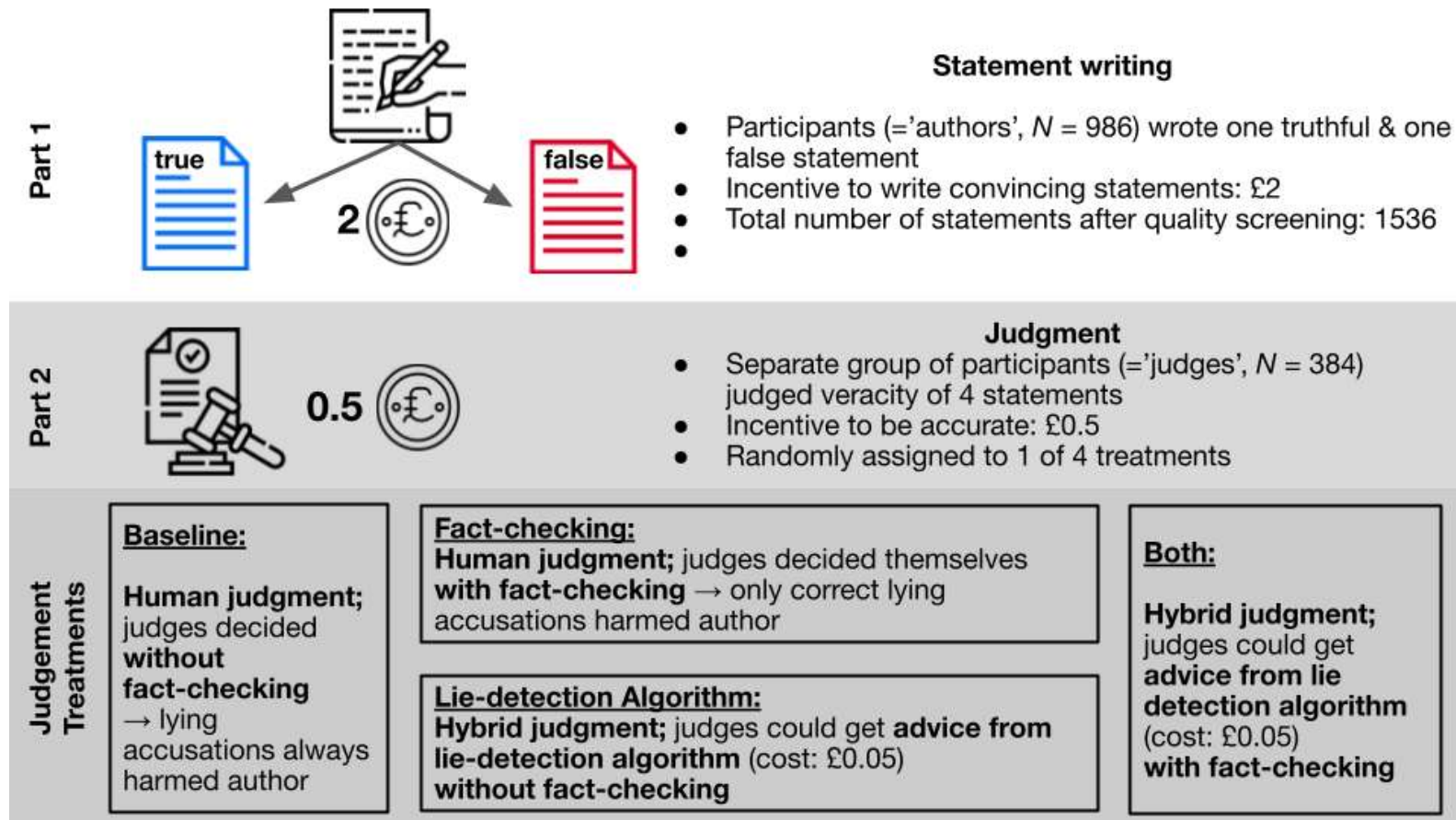

**Figure 1 | Overview of the study design.** The study consisted of two parts. In Part 1, participants (=authors) wrote one true and one false statement; In Part 2, a separate sample of participants (=judges) judged four statements. Four different treatments existed. In the 'Baseline', participants decided by themselves, without fact-checking (=all accusations led to a reduction of payoffs for the author); In the 'Fact-checking' treatment, participants decided by themselves, with fact-checking (only accusations that were justified led to a reduction of payoffs for the author); In the 'Lie-detection Algorithm' treatment, judges could purchase the advice from a lie-detection algorithm, yet fact-checking was not present. In the 'Both' treatment, judges could purchase advice from a lie-detection algorithm.

At the end of the study, all participants answered a series of questions measuring their beliefs about the accuracy of the algorithm (percentage of mistakes, percentage of false accusations, accuracy compared to the average human, accuracy compared to themselves). We informed those participants who did not have access to the lie-detection algorithm about the existence of an intelligent algorithm that was designed to predict the truthfulness of statements. This elicitation allowed us to assess whether the decision to use the algorithm correlated with subjective expectations about its performance.

## Results

**Algorithmic vs. human performance.** To empirically validate that our algorithm exceeded human performance at lie-detection, we compared human and algorithmic accuracy levels. Participants achieved a 50% accuracy rate, in line with previous findings documenting people's inability to discern truthful statements from lies (Verschuere et al. 2018). The lie-detection algorithm achieved a 67% accuracy rate, which significantly exceeds random guessing ($t = 7.95$, $p < .001$) and is comparable to previously developed lie-detection algorithms (Kleinberg and Verschuere 2021).

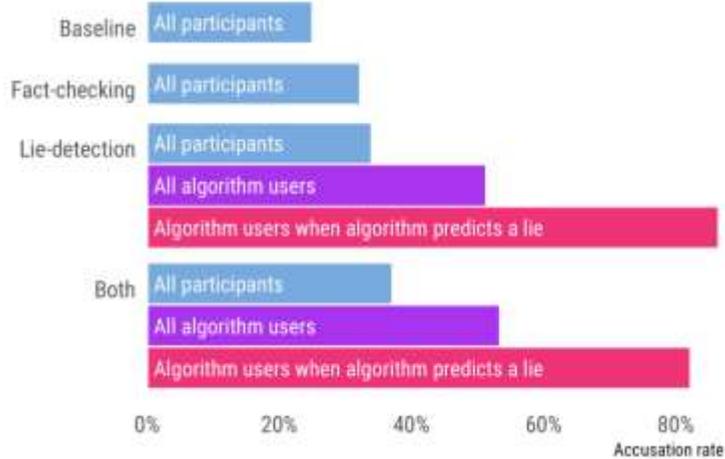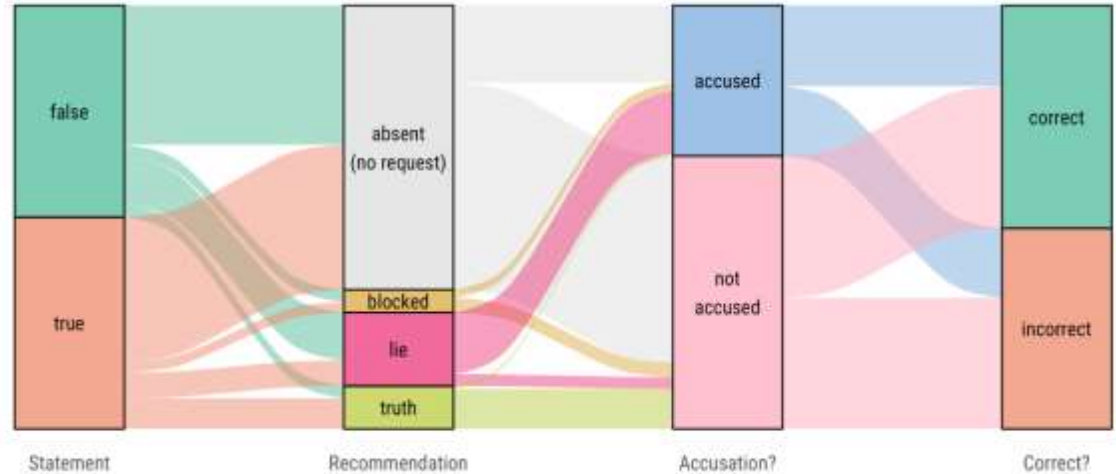

**Figure 2 | Main experimental results.** (A) Accusation rate in each of the four treatments. In the 'Lie-detection Algorithm' treatment and the 'Both' treatment (=where participants had access to our AI lie detection algorithm), accusation rates are also shown for the subset who elected to use the algorithm, as well as for the sub-subset of these participants whose lie-detection algorithm tagged the target statement as a lie. (B) Detailed stage-by-stage data for the 'Lie-detection Algorithm' treatment and the 'Both' treatment, showing the guess of the AI (not requested, blocked, lie truth) for false and true target statements, the subsequent decision of the participant depending on the algorithm's guess, and the accuracy of this decision.

**Accusation rates across treatments (Figure 2A).** In the 'Baseline' treatment (no fact-checking, no lie-detection algorithm), the accusation rate was 25%, even though participants knew for a fact that 50% of the statements were lies — which confirms the assumption that people typically refrain from accusing others of lying (Gilbert 1991, Köbis, Doležalová, et al. 2021). In the 'Fact-checking' treatment, we removed one possible reason for this reticence since false accusations did not harm accused players. Fact-checking increased accusation rates by 7 points, up to 32% ($t = -2.48$, $p = .014$). This finding suggests that participants were indeed sensitive to the risk of harming others through false accusations, even though the corresponding increase in accusation rates is not very large.

In the 'Lie-detection Algorithm' treatment (no fact-checking, lie-detection algorithm available), the availability of the algorithm had a comparable effect to the presence of fact-checking, increasing accusation rates to 34% ($t = 2.93$, $p = .004$). Still, while the rate at which a lie remains undetected reduces from 77% in the Baseline to 64% in the Lie-detection Algorithm treatment ($p = .005$ from regressions with clustered standard errors), the likelihood of false accusations increases only slightly and not at a statistically significant level (27% vs. 32%, $p = .188$). The overall accusation rate includes the decisions of participants who declined to use the algorithm, as well as the decisions of participants whose algorithm judged the target statement as truthful. If we only include the decisions of participants who elected to use the algorithm, the accusation rate increases to 53%. Furthermore, if we only include the decisions of participants whose algorithm flagged the target statement as a lie, the accusation rate drastically increases to 86%.

When comparing participants who use the algorithm with those who do not — either because they are in the 'Baseline' or because they choose not to purchase the prediction — we observe a significant increase in the false accusation rate (27% vs. 43%, $p = .035$).

Nevertheless, on the positive side, we observe a strong improvement in the correct detection of lies. The decrease in the likelihood of a lie going through as truth is more than twice as large as the increase in false accusation rates (75% vs. 38%, $p < .001$). In the simultaneous presence of fact-checking and the lie-detection algorithm (see 'Both' treatment in Figure 2A), general accusation rates, as well as the likelihood of a lie remaining undetected and false accusation rates, do not differ significantly ($p > .301$).

**Algorithmic uptake (Figure 2B).** Although individuals are much more likely to accuse others of lying when advised to do so by an algorithm, Figure 2B illustrates that introducing the lie-detection algorithm does not cause large changes at the collective level. The figure combines data from the two treatments in which the algorithm is available ('Lie-detection Algorithm' & 'Both'). As shown on the right-hand side, accuracy (53%) does not differ much from chance at the collective level.

Moving one step to the left, we can see that accusation rates (35%) are still low, considering that participants knew that 50% of the statements were lies. Moving one step further to the left elucidates why the collective changes are small: Only a minority of participants elect to use the algorithm at all (33% overall, not significantly different in the presence or absence of fact-checking, 31% vs. 35%, $p = .43$), which is not enough to observe large collective effects, either on accusation rates or on accuracy.

Further, authors preventing algorithmic judgment of their statements was also not an obstacle as only 16% chose to pay for blocking the application of the lie-detection algorithm to their two statements. This finding suggests that for lie-detection algorithms to bring substantial social change, their uptake must also be substantial. Accordingly, it is important to investigate individual predictors of the decision to use lie-detection algorithms to anticipate the magnitude of the social changes they might provoke.

There are arguably many such potential predictors, but we start here with the most obvious one: subjective confidence in the accuracy of the algorithm, both in absolute and relative terms. We collected four measures of this subjective confidence among participants (whether or not they had access to the algorithm or whether or not they ultimately decided to use it): two subjective judgments about the absolute performance of the algorithm (expected percentage of mistakes, and expected percentage of false accusations), and two subjective judgments in its relative performance (expected accuracy compared to the human average, and expected accuracy compared to one's own accuracy).

Table 1 displays the results of regression models predicting the frequency at which participants elect to use the algorithm as a function of their four judgments about its expected performance (restricted to the 'Lie-detection Algorithm' treatment). In the absence of fact-checking, all four judgments correlate with the decision to use the algorithm: whatever the measure of expected performance, higher expected performance is associated with greater algorithmic uptake. One interpretation of this finding is simply that people want value for their money. That is, they are less likely to purchase the algorithmic prediction (even for a meager cost of £0.05) when they have doubts about the performance of the algorithm.

**Table 1** | Regression of the frequency of requesting a hint in the 'Lie detection Algorithm' treatment on the beliefs about the general accuracy of the algorithm (in %), its false accusation rate (in %), its performance compared to an average human (from -5, the algorithm is better, to +5, human is better), and its performance compared to the participant (from -5, the algorithm is better, to +5, oneself is better). Standard errors are reported in parentheses. Significance coding: * $p < .10$, ** $p < .05$, *** $p < .01$

|  | (1) | (2) | (3) | (4) | (5) |
|---|---|---|---|---|---|
| Belief Accuracy | 0.00533*** (0.00157) |  |  |  | 0.00346* (0.00185) |
| Belief False Accusation |  | -0.00512*** (0.00156) |  |  | -0.00345** (0.00169) |
| Belief Algo vs. Average |  |  | -0.0374*** (0.0135) |  | -0.00756 (0.0197) |
| Belief Algo vs. Own |  |  |  | -0.0350** (0.0140) | -0.00502 (0.0194) |
| Constant | -0.00598 (0.0987) | 0.527*** (0.0756) | 0.315*** (0.0358) | 0.297*** (0.0362) | 0.252* (0.144) |
| N | 95 | 95 | 95 | 95 | 95 |
| R² | 0.111 | 0.103 | 0.077 | 0.063 | 0.166 |

However, a problem with this interpretation comes from the results of the same regressions when fact-checking is also present (i.e, in the 'Both' treatment). In this treatment (Table 2), all associations become non-significant: participants put a much lower weight on the performance of the algorithm once false accusations no longer hurt the accused. Overall, these results suggest that people do not think of the performance of the algorithm in terms of strategic gains but as a guarantee that it will not unduly harm others by incorrectly flagging their statements as lies.

**Table 2 |** Regression of the frequency of requesting a hint in the 'Both' treatment on the beliefs about the general accuracy of the algorithm (in %), its false accusation rate (in %), its performance compared to an average human (from -5, the algorithm is better, to +5, human is better), and its performance compared to the participant (from -5, the algorithm is better, to +5, oneself is better). Standard errors are reported in parentheses. Significance coding: * $p < .10$, ** $p < .05$, *** $p < .01$

|  | (1) | (2) | (3) | (4) | (5) |
|---|---|---|---|---|---|
| Belief Accuracy | 0.00184 (0.00211) |  |  |  | 0.00322 (0.00241) |
| Belief False Accusation |  | 0.00239 (0.00206) |  |  | 0.00380 (0.00241) |
| Belief Algo vs. Average |  |  | -0.000484 (0.0166) |  | 0.00524 (0.0234) |
| Belief Algo vs. Own |  |  |  | -0.0102 (0.0156) | -0.0141 (0.0197) |
| Constant | 0.243* (0.129) | 0.243** (0.0999) | 0.350*** (0.0389) | 0.350*** (0.0385) | -0.00859 (0.202) |
| N | 95 | 95 | 95 | 95 | 95 |
| R² | 0.008 | 0.014 | 0.000 | 0.005 | 0.043 |

## Discussion

Today, we live in a world where false accusations of lying come with costs, both to the accuser and the accused. One important feature of this current social equilibrium is that people generally refrain from accusing others of lying. We are interested in pre-emptively understanding the potential impact of novel AI-driven lie-detection and fact-checking technologies on this social equilibrium.

Specifically, we hypothesized that technologies that lower the probability or the cost of false accusations might disrupt the current social equilibrium and increase the rate at which people accuse each other of lying. We simulated different (technological) versions of the future in an incentivized experiment: futures in which accusations are systematically fact-checked, futures in which people have access to a lie-detection algorithm (which we prototyped for our experiment), and a combination of both.

At the collective level, we observed that accusation rates modestly increased in all experimental simulations of these technological futures. First, in an environment with fact-checking, where calling someone a liar does not always bear consequences but is verified first, people perceive accusations as less morally questionable. Here, they are significantly more likely to make such accusations than in an environment without fact-checking, where the perceived (moral) costs are higher.

Second, making a lie-detection algorithm available offers the opportunity to transfer the accountability for accusations from oneself to the system (Hohenstein and Jung 2020, Köbis, Bonnefon, et al. 2021). However, when participants can use an algorithm for lie detection, they only rely on its recommendations when they believe it makes accurate predictions. This finding suggests that in a morally controversial domain such as lie detection, algorithmic uptake is not purely driven by blame-shifting motives but also by the desire to rely on algorithmic support to make more accurate and fair judgments. Delegating a decision involving as much as calling someone a liar to an algorithm without a secure fact-checking process at least invokes considerations about the predictive power and reliability of such systems.

Our most striking result was that accusation rates climbed above 80% for people who used a lie-detection algorithm that flagged a statement as a lie. Accordingly, lie detection algorithms could have a strong disruptive potential for our current social equilibrium. Yet, in our experiment, low uptake of the algorithm weakened the disruptive impact — only 30% of participants elected to use the algorithm. But if algorithm uptake increases, our society might undergo significant transformations, for better or worse.

High accusation rates may strain our social fabric by fostering generalized distrust and further increasing polarization between groups that already find it difficult to trust one another. However, making accusations easier, especially if these accusations are reasonably accurate, may also lead to beneficial effects by discouraging insincerity and promoting truthfulness in personal and organizational communications. Accuracy is an important factor here: we know that individuals can easily get false confidence in their ability to detect lies. Such is the case when they are exposed to pseudo-scientific methods of spotting liars (e.g., after learning the techniques of the TV show "Lie to me" Levine et al. 2010). An advantage of lie-detection algorithms is that they can be properly tested and certified for above-human accuracy in a specific domain (Guszcza et al. 2018).

Estimating the positive and negative social effects of lie-detection algorithms is not easy in a lab experiment since these effects may unfold slowly, in a cumulative manner, over a long period. Lab experiments are not the best tool for estimating these long-term cumulative effects, which is one limitation of our current work. But even if we cannot fully assess the magnitude and probability of these social changes, it seems reasonable to accept that to maintain a positive balance between benefits and costs, we will need to be mindful of the performance of lie-detection algorithms before making them massively available, and to use them responsibly as individuals and organizations, taking into account their limitations.

Our findings provide at least an encouraging signal in that direction: at least in one version of the future (lie-detection algorithms without systematic fact-checking of accusations), algorithmic uptake depends on the perceived accuracy of algorithms and, in particular, their false accusation rates. This finding suggests that individuals may be mindful of the performance of lie-detection algorithms and use them somewhat responsibly to make accusations.

Organizations, on the other hand, may not always be so careful. Some managerial domains, such as negotiations with suppliers or clients, might be early adopters of lie-detection algorithms and pressure other domains, such as human resources, to do the same. Since suspicion about out-groups may be more socially acceptable than suspicion within the in-group, using lie-detection algorithms when dealing with other organizations may pave the way for their use within an organization. Behavioral science has a crucial role to play in anticipating these dynamics and carefully managing the transition to a high-accusation social world.

# Supplementary Material

**Inclusion criteria for Statements**

The research assistants checked whether the authors wrote a meaningful statement about their activities (or, for the false statements, as if they were going to carry it out) as intended. For the truthful statements, they further verified that the additional question asking for supportive information fitted and reinforced the participant's entry. The third criterion was automatically applied and flagged all statements with less than 150 characters. If at least one statement of the authors failed at least one verification, we took out this author completely and did not use any of his/her statements for Part 2.

# Experimental Instructions - Authors

## Welcome

Welcome to this online experiment and thank you for your participation!

We will now give you detailed instructions. **Please read them carefully.**

You receive a base payment of £1.00 that will be paid out today.
In addition, you can earn extra money in the experiment depending on your choices.
You will receive this additional payment at a later date.

## Introduction

### Overview

In this experiment, you will write a **short statement about your most significant non-work-related activity** in the next seven days.
This statement must be **truthful** and must describe an activity you truly will carry out.

We will show your statement to a future participant.
This participant will **guess whether it is truthful or false.**
This future participant receives a financial bonus for correctly classifying lies as lies and truthful statements as truthful.

### Payment

For your participation in this study, you receive a base payment of £1.00 that will be paid out today.
On top of that you can earn an additional payment of up to £4.00.
**The additional payment depends on the judgment of a future participant.**

Namely, if a future participant believes your statement to be true, you earn an additional £2.00.
If a future participant believes your statement to be false, you earn an additional £2.00 only with 50% probability.

You will receive the additional payment (if any) after the future participant has taken part in the study.

## Your Activity

Please describe your most significant, non-work-related activity in the next seven days.
Your statement must be **truthful** and must describe an activity you truly will carry out.
This activity should be specific, have a clear start and an end time, and not be a continuous or daily activity.
Do not reveal any personally identifiable information such as names or addresses.
Furthermore, do not include any potentially harmful or illegal content (incl., hate speech,

etc.).
Your statement should be about 300-400 characters.

Name for your activity:

[                                                                        ]

Please describe your activity as specific as possible:

[                                                                        ]

Which information can you give us to reassure us that you are telling the truth? (1-2 sentences)

[                                                                        ]

The experimenters will check whether you have written a meaningful, formulated text.
If not, you will not receive any payment.

**Activities of Other Participants**

We further would like to know which activities you will NOT carry out.
Below you find three activities of other participants of this study.
Please indicate which of these activities do not apply to you and you will not carry out.

☐[Other Activity 1]

☐[Other Activity 2]

☐[Other Activity 3]

**Activities of Other Participants**

Second Statement

You indicated that you will not carry out the activity "[Other Activity 1]".
For this activity, we ask you to make another statement.
This other statement must be false.
Please pretend that this would be your most important activity for the next week.
Provide a made-up statement similar to the one about your truthfully reported activity.

Payment

For this statement you can earn £4.00, as well.
The additional payment again depends on the judgments of a future participant.

Namely, if a future participant believes your statement to be true, you earn an additional £2.00.
If a future participant detects your statement as a lie, you receive no additional payment.

You will receive the additional payment (if any) after the future participant has taken part in the study.

**Your Made-up Statement**

You indicated that you will not carry out the activity "[Other Activity 1]".
Please pretend that this would be your most important activity for the next week.
Provide a made-up statement similar to the one about your truthfully reported activity.
This statement should be about 300-400 characters.

Name for the activity:
[Other Activity 1]

Made-up description of your activity:

|  |
|--|
|  |

The experimenters will check whether you have written a meaningful, formulated text.
If not, you will not receive any payment.

**Lie Detection Algorithm** *[shown to half of all authors]*

Recall that future participants will judge the truthfulness of your statements.
**For each statement a future participant believes to be true, you earn an additional £2.00.**
The future participants will be able to use a state-of-the-art **artificially intelligent lie detection algorithm** for their judgments.
This algorithm can analyze text and make predictions about the truthfulness of the content.
The future participants can pay a small fee to obtain an algorithmic prediction of the truthfulness of your statement.
You can prevent this use by paying a small fee of £0.30.
This will block the application of the lie detection algorithm to *both* of your statements.

Do you want to prevent the use of the lie detection algorithm for your statements?

☐ Yes
☐ No

**Results**

The main part of the experiment is now finished.
We will transfer your base payment of £1.00 as soon as possible.

You receive an additional £2.00 for each of your statements that a future participant believes to be true.
We will pay out the additional payment (if any) after the future participants have taken part in the study.
Note that this may take some time.

**Assessments**

Besides the judgment of future participants, a state-of-the-art intelligent algorithm was designed to predict truthfulness.
This algorithm can analyze text and make predictions about the truthfulness of the content.
Before the end of the experiment, we would like to ask for your opinion in some short questions.
These questions concern the performance of this algorithm.

Question 1

How frequently do you think the state-of-the-art intelligent algorithm **correctly predicts** whether a statement is true or false?

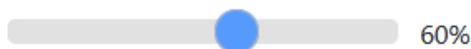

Question 2

How good do you think the **average human performance** is compared to the performance of the intelligent lie detection algorithm in predicting whether a statement is true or false?

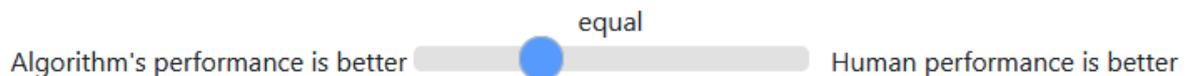

Question 3

How frequently do you think the intelligent algorithm **incorrectly predicts a lie** although it is actually a true statement?

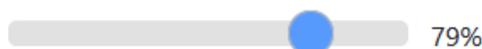

Question 4

How much confidence do you have in your assessment of the performance of the algorithm?

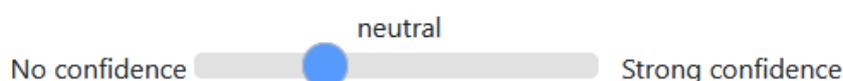

**Survey**

Please fill out this final survey before finishing the experiment.

How old are you?

<div style="border:1px solid black; width:200px; height:30px;"></div>

What is your gender?

☐ female
☐ male
☐ other/non-binary

What is your highest educational degree?

☐ No degree
☐ High school
☐ Bachelor
☐ Master
☐ PhD

If you go/went to university, what is/was your major?

☐ Not applicable
☐ Economics
☐ Law
☐ Psychology
☐ Political sciences
☐ Medicine
☐ Natural sciences
☐ Engineering
☐ Other social sciences
☐ Other

What is your employment status?

☐ Unemployed
☐ Part-time
☐ Full-time

How familiar are you with new technologies such as machine learning?

☐ Not familiar at all
☐ Rather not familiar
☐ Neutral
☐ A little familiar
☐ Very familiar

**End of Experiment**

The experiment is now finished.
Please click on the button below to return to Prolific.
You can only receive payment after being redirected to Prolific.
You will receive your payment as soon as possible.

Back to Prolific

# Experimental Instructions - Judges

## Welcome

Welcome to this online experiment and thank you for your participation!
We will now give you detailed instructions. **Please read them carefully.**
You receive a base payment of £1.00 that will be paid out today.
In addition, you can earn extra money in the experiment depending on your choices.
You will receive this additional payment at a later date.

## Introduction

Overview

In this experiment, you will read short statements about a non-work-related activity that are either truthful or lies.
Past participants of this study made these statements.
In total, we show you 4 different statements.
**For each statement, you need to guess whether it is truthful or a lie.**
Half of *all* statements are truthful and half of them are lies.
However, the actual number of truthful and lied statements we show you is random.

Payment

For your participation, you receive a base payment of £1.00.
**On top of that you receive a bonus for each correct guess.**
That is, if you judge a truthful statement to be truthful, you receive £0.50.
Likewise, if you judge a lied statement to be a lie, you receive £0.50.
Otherwise, you receive no bonus.

Lie detection algorithm *[Lie-detection treatments]*

You can use a state-of-the-art **artificially intelligent lie detection algorithm** for your judgments.
You can pay a small fee of £0.05 to obtain an algorithmic prediction of the truthfulness of the statement.
For some statements, the prediction of the algorithm is not available.
In this case, you would not be charged the £0.05 for purchasing a prediction.

Consequences of lie accusations for the author of the statement

*[Baseline without fact-checking]*
**Whenever you accuse an author of lying, this author is punished.**
**In this case, this author loses 40% of his/her total achievable payoff.**
**This happens *regardless* of whether you are correct in accusing the author of lying or not.**

*[Treatment with fact-checking]*
**Whenever you *correctly* accuse an author of lying, this author is punished.**

**In this case, this author loses 40% of his/her total achievable payoff.
If you *incorrectly* accuse an author of lying, this author is *not* punished.**

**Statement No. 1**

On the next screen, we will show the 1st statement we ask you to judge.
Please click Next to continue.

**Judgment**

**Name of the activity:**
[Activity 1]

**Statement and description of the activity:**
[Description of activity 1]

Do you want to purchase a prediction of the state-of-the-art algorithm for this statement for £0.05?
For some statements, the prediction of the algorithm is not available.
In this case, you would not be charged the £0.05 for purchasing a prediction.

[ Purchase for £0.05 ]

The algorithm predicted this statement to be **a lie / truthful**.

What do you think:
Is this statement truthful or a lie?

[ Truth ]   [ Lie ]

**Results**

*[Treatments without lie-detection algorithm]*
You rated 1 out of 4 statements correctly.
Therefore, you receive a payoff of £0.50 on top of your base payment of £1.00.
Your total payoff is thus **£1.50**.
We will transfer your total payoff as soon as possible.

*[Lie-detection treatments]*
You rated 3 out of 4 statements correctly.
You therefore earned £1.50.

You requested 2 predictions from the algorithm for £0.05 each.
For 0 of your requested hints, the prediction was not available.
The remaining 2 hints cost you £0.10 in total.

Therefore, you receive a payoff of £1.50 – £0.10 = £1.40 on top of your base payment of £1.00.
Your total payoff is thus **£2.40**.
We will transfer your total payoff as soon as possible.

**Assessments**

*[Treatments without lie-detection algorithm]*
Besides your judgment of the statements, a state-of-the-art intelligent algorithm was designed to predict truthfulness.
This algorithm can analyze text and make predictions about the truthfulness of the content.
Before the end of the experiment, we would like to ask for your opinion in some short questions.
These questions concern the performance of this algorithm.

*[Lie-detection treatments]*
Before the end of the experiment, we would like to ask for your opinion in some short questions.

Question 1

How frequently do you think the state-of-the-art intelligent algorithm correctly predicts whether a statement is true or false?

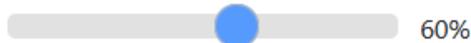

Question 2

How good do you think the average human performance is compared to the performance of the intelligent lie detection algorithm in predicting whether a statement is true or false?

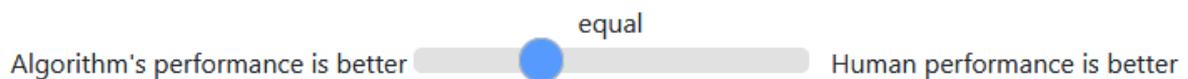

Question 3

How good do you think **your performance** is compared to the performance of the intelligent lie detection algorithm in distinguishing truth from lies?

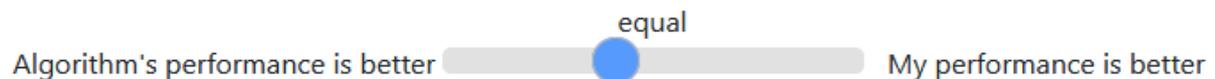

Question 4

How frequently do you think the intelligent algorithm incorrectly predicts a lie although it is actually a true statement?

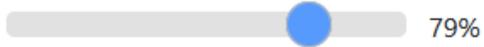 79%

Question 5

How much confidence do you have in your assessment of the performance of the algorithm?

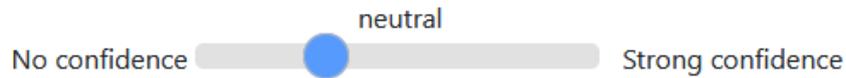

**Survey**

Please fill out this final survey before finishing the experiment.

How old are you?

What is your gender?

☐ female
☐ male
☐ other/non-binary

What is your highest educational degree?

☐ No degree
☐ High school
☐ Bachelor
☐ Master
☐ PhD

If you go/went to university, what is/was your major?

☐ Not applicable
☐ Economics
☐ Law
☐ Psychology
☐ Political sciences
☐ Medicine
☐ Natural sciences
☐ Engineering
☐ Other social sciences
☐ Other

What is your employment status?

☐ Unemployed

☐Part-time
☐Full-time

How familiar are you with new technologies such as machine learning?

☐Not familiar at all
☐Rather not familiar
☐Neutral
☐A little familiar
☐Very familiar

**End of Experiment**

The experiment is now finished.
Please click on the button below to return to Prolific.
You can only receive payment after being redirected to Prolific.
You will receive your payment as soon as possible.

Back to Prolific